 \documentclass[aip,apl,a4paper, reprint,amsmath,amssymb,floatfix]{revtex4-1}

 \usepackage{amsmath}
 \usepackage{bm}
 \usepackage{graphicx}
 \usepackage{SIunits}

\begin{document}

 \title{Extraction of the $\beta$-factor for single quantum dots coupled to a photonic crystal waveguide}
 \date{\today}
 \author{Henri Thyrrestrup}
 \email{htni@fotonik.dtu.dk}
 \author{Luca Sapienza}
 \author{Peter Lodahl}
 \email{pelo@fotonik.dtu.dk}
 \affiliation{DTU Fotonik, Department of Photonics Engineering, Technical University of Denmark, Building 345V, DK-2800 Kgs. Lyngby, Denmark}

        % Edit this

 \begin{abstract}
      We present measurements of the $\beta$-factor, describing the coupling efficiency of light emitted by single InAs/GaAs semiconductor quantum dots into a photonic crystal waveguide mode. The $\beta$-factor is evaluated by means of time-resolved frequency-dependent photoluminescence spectroscopy. The emission wavelength of single quantum dots is temperature tuned across the band edge of a photonic crystal waveguide and the spontaneous emission rate is recorded. Decay rates up to $\unit{5.7}{\nano\reciprocal\second}$, corresponding to a Purcell factor of 5.2, are measured and $\beta$-factors up to 85\% are extracted. These results prove the potential of photonic crystal waveguides in the realization of on-chip single-photon sources.
 \end{abstract}

 \maketitle

     The construction of efficient single-photon sources  is a research field that has attracted a lot of interest, in particular in the scope of applications in quantum cryptography and quantum information technology\cite{Scheel2009Singlephoton}. A fundamental parameter characterizing the quality of a single-photon source is the $\beta$-factor that describes the efficiency of the emission into a single photonic mode. Maximizing the $\beta$-factor is a pivotal element in implementing an efficient single-photon source\cite{Barnes2002Solidstate}. Semiconductor quantum dots are very promising solid-state single-photon sources\cite{Michler2000Quantum,Shields2007Semiconductor} thanks to their quantized energy levels, high optical quantum efficiency\cite{Johansen2008Size} and coherent emission properties\cite{Santori2002Indistinguishable}. The harvesting of the emitted photons into a single mode is typically achieved by placing the emitters within a photonic nanostructure, such as a resonant cavity\cite{Vahala2003Optical}. By coupling a quantum dot to a photonic nanostructure the spontaneous emission rate increases, through the so-called Purcell effect. This results in a reduction of the emission time-jitter and improves the coherence properties of the single-photon source. Micropillar cavities\cite{Heindel2010Electrically} and photonic crystal (PhC) cavities\cite{Englund2005Controlling,Kress2005Manipulation} for example, are characterized by high-$Q$ factors and small mode volumes and allow to reach high $\beta$-factors. Nevertheless, they operate in a narrow bandwidth and the geometry implies that light is predominantly emitted in the out-of-plane direction. Both these limitations reduce the possible implementation of such photonic nanostructures into large scale on-chip devices.

     As theoretically proposed\cite{Lecamp2007Very,Rao2007Single} and experimentally demonstrated\cite{Hansen2008Experimental,Dewhurst2010Slowlightenhanced}, PhC waveguides represent a promising alternative thanks to the efficient broadband coupling of the single-photon emission into a propagating mode, which is enhanced by slow light.  So far the $\beta$-factor in PhC waveguides has been determined by comparing the decay rate of individual quantum dots to the average decay rate of a quantum dot ensemble\cite{Hansen2008Experimental}. As the position and dipole orientation of the quantum dots relative to the local electric field are not controlled, such a procedure provides a statistical measure of the $\beta$-factor. In the present work, direct measurements are presented: a single quantum dot is temperature tuned in the vicinity of the waveguide band edge and the radiative decay rate is measured at various frequencies. This allows us to eliminate the statistical average in the position and dipole-orientation of the quantum dots and directly extract the $\beta$-factor of individual quantum dots.

    \begin{figure}[bp]
    \centering
    \includegraphics[]{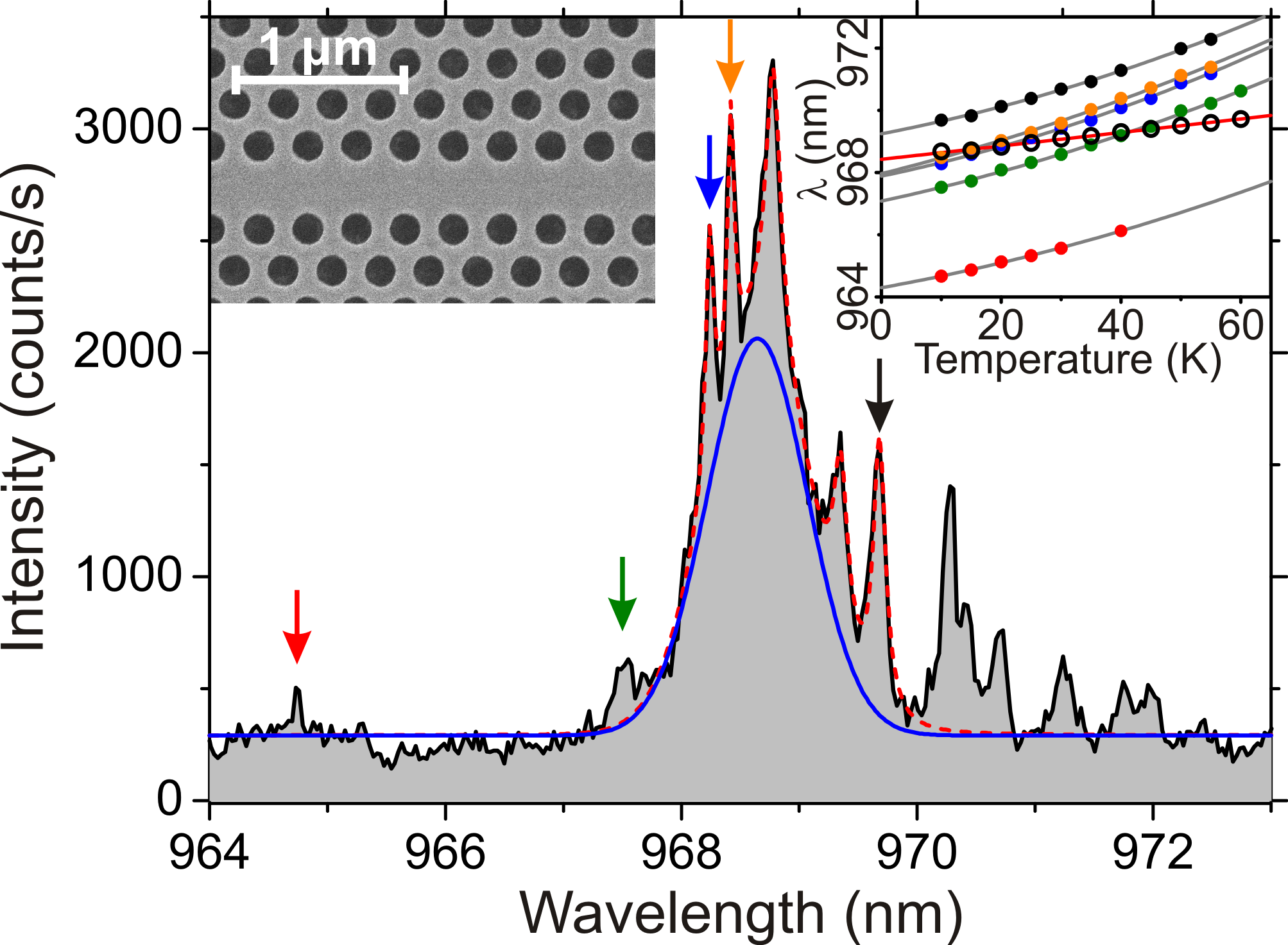}
    \caption{Photoluminescence spectrum (black line) of InAs quantum dots embedded in a PhC waveguide measured at $10\kelvin$ with a pump power of \unit{\sim 1}{\micro\watt}. The dashed red line shows a multi-Lorentzian fit of the sharp quantum dot lines and a Gaussian function to fit the broad peak (blue line), which is a signature of the PhC crystal waveguide band edge. (Top left inset) SEM image of the photonic crystal waveguide. (Top right inset) Temperature dependence of the quantum dot emission wavelength of 5 selected quantum dots, marked with arrows in the main panel (filled circles), and of the PhC waveguide band edge (open circles). The lines are second order polynomial fits to the data.}
    \label{fig:Tempdepend}
    \end{figure}

    The sample under study consists of a \unit{150}{\nano\meter} thick GaAs PhC membrane with lattice constant $a=\unit{256}{\nano\meter}$ and hole radius $r = 0.30a$, containing a single layer of InAs quantum dots with a density of \unit{~250}{\micro\rpsquare\meter} in the center. The waveguide is formed by leaving out a single row of holes in the PhC structure and it has a length of \unit{100}{\micro\meter}. A scanning electron microscope (SEM) image of the waveguide is shown in Fig.~\ref{fig:Tempdepend}. The sample is positioned in a helium flow cryostat and the optical characterization is carried out in a confocal configuration. The quantum dots are excited with a Ti:Sapphire pulsed laser emitting at a wavelength of \unit{850}{\nano\meter} (with \unit{2}{\pico\second} pulses and a repetition rate of \unit{76}{\mega\hertz}), corresponding to excitation of carriers in the wetting layer. The excitation beam is focused with a NA=0.60 objective and the same objective is used to collect the emitted photons. A spatial filter results in a collection spot size diameter (FWHM) of \unit{1.4}{\micro\meter}. The collected photons are then dispersed in a \unit{0.67}{\meter} spectrograph with a resolution of \unit{0.15}{\nano\meter} and sent to a silicon avalanche photo diode (APD) with a temporal resolution of \unit{280}{\pico\second}.

    An example of a photoluminescence (PL) spectrum recorded at a temperature of \unit{10}{\kelvin} is shown in Fig.~\ref{fig:Tempdepend}. Several narrow peaks, due to the emission of single quantum dots are visible on top of a broader peak that is the spectral signature of the band edge of the PhC waveguide. The spectral position of the broad peak is at \unit{968.7}{\nano\meter}, which is in very good agreement with the band edge position (\unit{968.4}{\nano\meter}) obtained by a 3D band structure calculation (MPB\cite{Johnson2001:mpb} package) using  $n=3.44$ for the refractive index of GaAs. In contrast, sharp high intensity spectral resonances have been observed near the band edge of PhC waveguides resulting from disorder induced  Anderson localization modes \cite{Sapienza2010Cavity}. We note that Anderson localization is not observed in the present sample presumably due to enhanced out-of-plane scattering. From SEM images of the cleaved samples indeed surface roughness is observed on the bottom side of the membranes.

    The photonic nature of the broad peak is further substantiated by the temperature dependence of its spectral position. When changing the temperature of the sample the quantum dot emission wavelength shifts due to the change in the semiconductor band gap, whereas the photonic modes are affected by the change in the refractive index of the GaAs membrane. The emission wavelength of 5 selected quantum dots and of the waveguide band edge is plotted in Fig.~\ref{fig:Tempdepend} as a function of temperature. The temperature shift of the quantum dots ($\approx\unit{0.05}{\nano\meter\per\kelvin}$) is consistent with previously reported values\cite{Kiraz2001Cavityquantum}, and is larger than that of the waveguide mode ($\approx\unit{0.02}{\nano\meter\per\kelvin}$). This relative difference in temperature dependence can be used to tune the quantum dots into resonance with the waveguide band edge.

    \begin{figure}[tbp]
    \centering
    \includegraphics[]{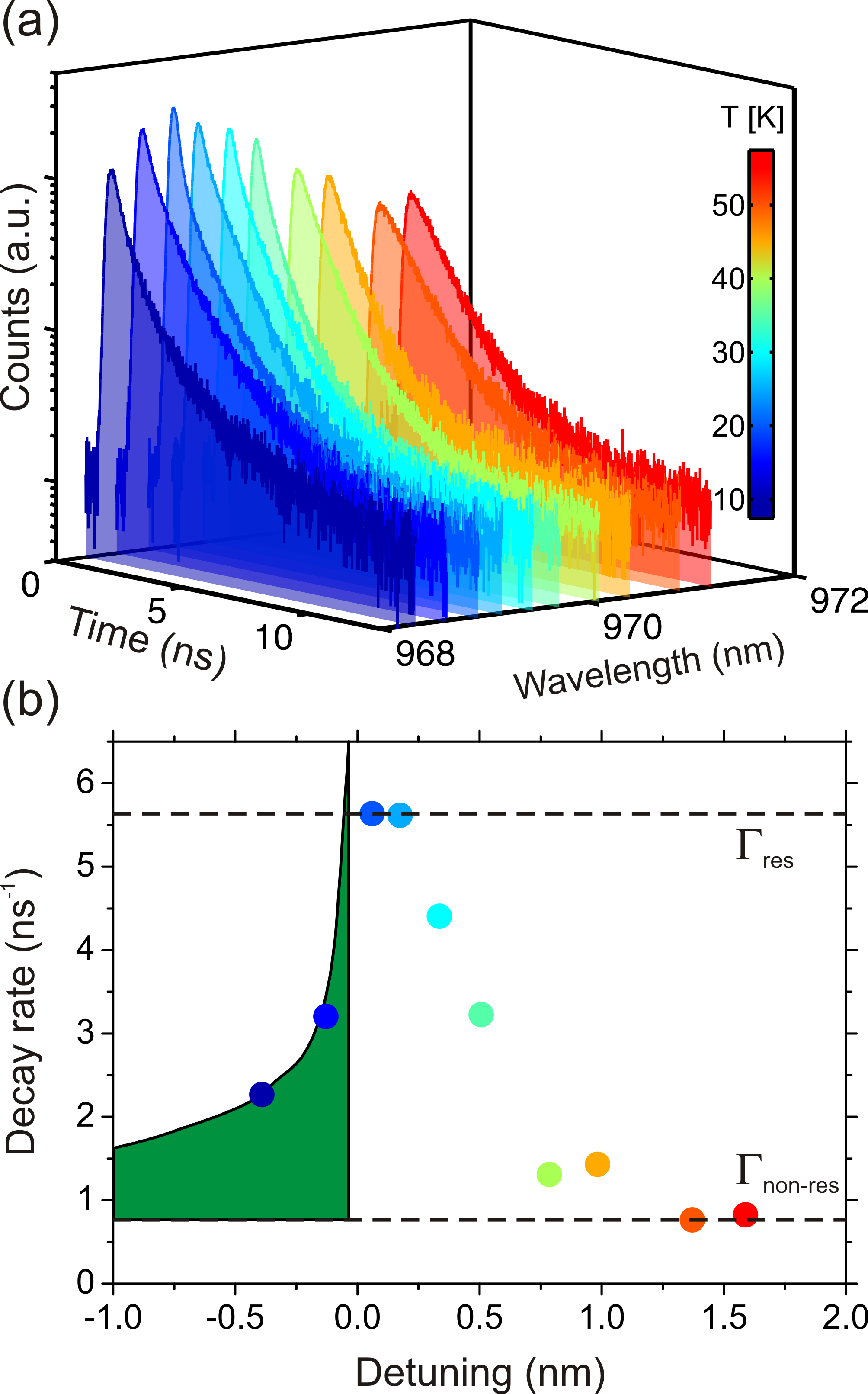}
    \caption{(a) Decay curves of a single quantum dot (QD3) measured with \unit{5}{\kelvin} steps in a temperature range between \unit{10}{\kelvin} and \unit{60}{\kelvin} and plotted as a function of the emission wavelength. (b) Decay rates of QD3 extracted from the data shown in panel (a) as a function of detuning relative to the waveguide band edge (filled circles, whose color relates to the temperature colorbar in panel (a)). The two dashed lines labeled with $\Gamma_{\text{res}}$ and $\Gamma_{\text{non-res}}$ mark the fastest decay rate on resonance with the PhC waveguide band edge and the slowest decay rate when the quantum dot emission lies in the PhC band gap, respectively. The solid line represents the decay rate calculated for a lossless PhC waveguide where the out-of-plane radiation contribution is set to $\Gamma_{\text{non-res}}$. The curve has been scaled by a factor $0.4$ to match the experimental data points accounting for spatial mismatch between the quantum dot position and polarization relative to the waveguide electric field.}
    \label{fig:DecayCurves}
    \end{figure}

    Examples of decay curves are shown in Fig.~\ref{fig:DecayCurves}(a) as a function of the emission wavelength, recorded at temperatures between \unit{10}{\kelvin} and \unit{60}{\kelvin}. The initial slope of the decay curves changes significantly with temperature and is steepest at \unit{15}{\kelvin}. The decay rates are extracted by fitting the curves with a bi-exponential model convoluted with the APD instrument response function and including the measured background level. The fastest of the two exponents corresponds to the total measured decay rate: $\Gamma_{\text{tot}}=\Gamma_{\text{wg}}+\Gamma_{\text{rad}}+\Gamma_{\text{non-rad}}$ that contains the radiative decay rate into the waveguide $\Gamma_{\text{wg}}$, out-of-plane radiation $\Gamma_{\text{rad}}$, and the non-radiative decay rate $\Gamma_{\text{non-rad}}$. The slow exponent contains contributions from fine structure effects\cite{Johansen2010Probing}.

    The extracted decay rates for QD3 are plotted in Fig.~\ref{fig:DecayCurves}(b) as a function of the detuning between the quantum dot and the PhC waveguide band edge. Going from negative towards zero detuning, the measured decay rate increases reaching a maximum value of $\Gamma_{\text{res}}=\unit{5.7}{\nano\reciprocal\second}$ on resonance. This corresponds to an enhancement of the spontaneous emission decay rate described by the Purcell factor $F_p=\Gamma_{\text{res}}/\Gamma_0$ of 5.2, where $\Gamma_0=\unit{1.1}{\nano\reciprocal\second}$ is the decay rate measured on a quantum dot in a homogenous medium. The Purcell factor is  $4$ times larger than previously observed for quantum dots coupled to a PhC waveguide\cite{Hansen2008Experimental}. For positive detunings, the measured decay rates decrease monotonically reaching a minimum value of $\Gamma_{\text{non-res}}=\unit{0.8}{\nano\reciprocal\second}$.

    These data can be used to extract the coupling efficiency of the emission from a single quantum dot into the waveguide mode, described by the $\beta$-factor:
    \begin{equation}
    \beta = \frac{\Gamma_{\text{wg}}}{\Gamma_{\text{tot}}}=\frac{\Gamma_{\text{res}}-\Gamma_{\text{non-res}}}{\Gamma_{\text{res}}}.\label{eq:beta}
    \end{equation}
    The decay rate $\Gamma_{\text{wg}}$ on resonance can be evaluated as $\Gamma_{\text{res}}-\Gamma_{\text{non-res}}$ when assuming $\Gamma_{\text{rad}}+\Gamma_{\text{non-rad}}$ to be constant in the considered wavelength range. From the measurements on QD3 we retrieve $\beta=85\%$ on resonance. Note that this corresponds to a lower bound of the real $\beta$-factor since tuning the quantum dots further away from the band edge is likely to reduce the coupling to the waveguide even further. A larger tuning range could be obtained by implementing alternative tuning schemes like electrical tuning\cite{Laucht2009Electrical} or gas tuning\cite{Mosor2005Scanning}. We note that in ref.~\onlinecite{Hansen2008Experimental} variations in $\Gamma_{\text{non-res}}$ between \unit{0.05-0.43}{\nano\reciprocal\second} were observed, which  would result in $\beta$-factors between $92\%-99\%$. The observed $\beta$-factors are competitive to numbers reported for quantum dots coupled to PhC cavities\cite{Chang2006Efficient,Kress2005Manipulation}.

    \begin{figure}[tbp]
    \centering
    \includegraphics[]{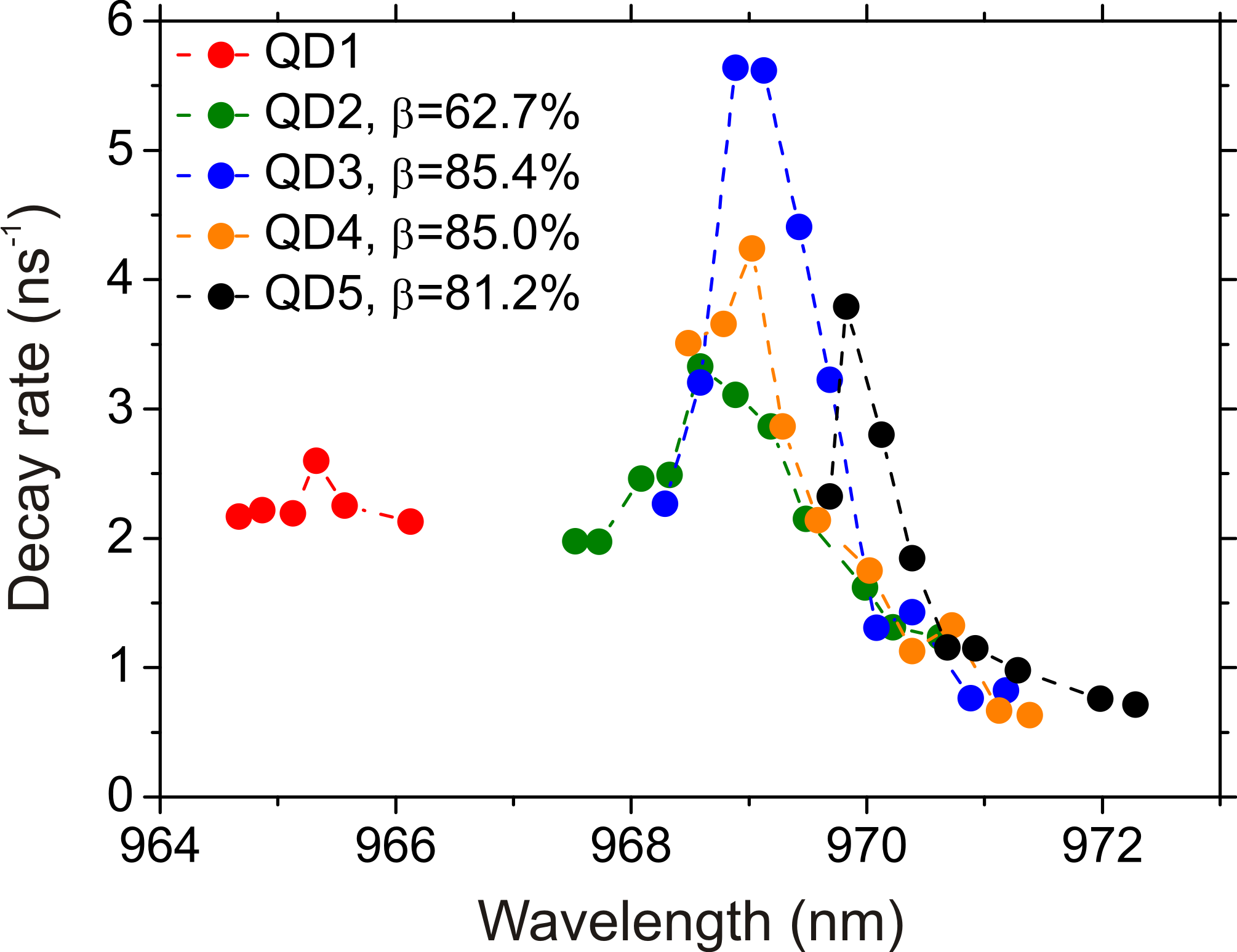}
    \caption{Decay rates of the 5 quantum dots marked with arrows in the spectrum in Fig.~\ref{fig:Tempdepend} plotted as function of the emission wavelength. The extracted $\beta$-factors for four quantum dots are shown in the legend.}
    \label{fig:DecayRates}
    \end{figure}

    A similar study has been carried out on four quantum dots positioned within the same collection spot. The extracted decay rates are shown in Fig.~\ref{fig:DecayRates} and they all show an enhancement around \unit{969}{\nano\meter} and a decline for longer wavelength indicating that the quantum dots couple to the waveguide mode. We extract $\beta$-factors between $63\%-85\%$, demonstrating that all the measured quantum dots couples efficiently to the waveguide.

    A quantum dot far detuned (approximately \unit{-2}{\nano\meter}) has also been studied: an almost constant decay rate of \unit{~2}{\nano\reciprocal\second} is observed throughout the tuning range. The flat dispersion is consistent with the quantum dot being coupled to the PhC waveguide and the large decay rate indicates a very efficient coupling of this quantum dot into the waveguide mode in a bandwidth larger than \unit{5}{\nano\meter}.

    Measuring the quantum dot decay rates at various detunings allows to directly map out the frequency dependence of the local density of states in the vicinity of the waveguide band edge. The decay rates of quantum dots coupled to a lossless PhC waveguide can be calculated from the simulated group velocity\cite{Rao2007Single}, and are shown by the solid line in Fig.~\ref{fig:DecayCurves} (b). Going from negative to zero detuning, the theoretical decay rate increases and is predicted to diverge near the waveguide band edge and drop abruptly for positive detuning inside the PhC band gap. Our experimental data show that the real local density of states is broadened and the divergence is removed, both these effects are signatures of absorption and scattering losses in the waveguide structure, as theoretical predicted in ref.~\onlinecite{Pedersen2008Limits}.

    In conclusion, we have demonstrated that the coupling efficiency of a single emitter into a PhC waveguide mode can be determined by measuring the decay rate of single semiconductor quantum dots tuned across the band edge. The efficiency of the coupling reaches values above 85\%, which proves the promising potential of photonic crystal waveguides for on-chip quantum information processing.

    We gratefully acknowledge S{\o}ren Stobbe for sample fabrication and the Council for Independent Research (Technology and Production Sciences and Natural Sciences) and the Villum Kann Rasmussen Foundation for financial support.

     %\nocite{*}  % Remove this!!
     %\bibliographystyle{aipnum4-1long}
     %\bibliographystyle{aipnum4-1}
     %\bibliographystyle{apl_custom}
     %\bibliography{articles}

 \end{document}